\documentclass[12pt]{article}
\usepackage{amsmath,amssymb,graphics,amsthm,psfrag,subfigure}

\newcommand{\ie}{{\em i.e.}}

\newcommand{\eg}{{\em e.g.}}

\newcommand{\secref}[1]{Section~\ref{#1}}
\newcommand{\figref}[1]{Fig.~\ref{#1}}

\newcommand{\thrmref}[1]{Theorem~\ref{#1}}

\newcommand{\corolref}[1]{Corollary~\ref{#1}}
\newcommand{\appref}[1]{Appendix~\ref{#1}}

\newtheorem{thrm}{Theorem}
\newtheorem{prop}{Proposition}

\newtheorem{corol}{Corollary}

\DeclareMathAlphabet{\mathbsf}{OT1}{cmss}{bx}{n}
\DeclareMathAlphabet{\mathssf}{OT1}{cmss}{m}{sl}
\DeclareMathAlphabet{\mathcsf}{OT1}{cmss}{sbc}{n}

\DeclareSymbolFont{bsfletters}{OT1}{cmss}{bx}{n}  
\DeclareSymbolFont{ssfletters}{OT1}{cmss}{m}{n}
\DeclareMathSymbol{\bsfGamma}{0}{bsfletters}{'000}
\DeclareMathSymbol{\ssfGamma}{0}{ssfletters}{'000}
\DeclareMathSymbol{\bsfDelta}{0}{bsfletters}{'001}
\DeclareMathSymbol{\ssfDelta}{0}{ssfletters}{'001}
\DeclareMathSymbol{\bsfTheta}{0}{bsfletters}{'002}
\DeclareMathSymbol{\ssfTheta}{0}{ssfletters}{'002}
\DeclareMathSymbol{\bsfLambda}{0}{bsfletters}{'003}
\DeclareMathSymbol{\ssfLambda}{0}{ssfletters}{'003}
\DeclareMathSymbol{\bsfXi}{0}{bsfletters}{'004}
\DeclareMathSymbol{\ssfXi}{0}{ssfletters}{'004}
\DeclareMathSymbol{\bsfPi}{0}{bsfletters}{'005}
\DeclareMathSymbol{\ssfPi}{0}{ssfletters}{'005}
\DeclareMathSymbol{\bsfSigma}{0}{bsfletters}{'006}
\DeclareMathSymbol{\ssfSigma}{0}{ssfletters}{'006}
\DeclareMathSymbol{\bsfUpsilon}{0}{bsfletters}{'007}
\DeclareMathSymbol{\ssfUpsilon}{0}{ssfletters}{'007}
\DeclareMathSymbol{\bsfPhi}{0}{bsfletters}{'010}
\DeclareMathSymbol{\ssfPhi}{0}{ssfletters}{'010}
\DeclareMathSymbol{\bsfPsi}{0}{bsfletters}{'011}
\DeclareMathSymbol{\ssfPsi}{0}{ssfletters}{'011}
\DeclareMathSymbol{\bsfOmega}{0}{bsfletters}{'012}
\DeclareMathSymbol{\ssfOmega}{0}{ssfletters}{'012}

\newcommand{\genericRV}[1]{\mathssf{#1}} 
\newcommand{\genericS}[1]{\mathbf{#1}} 
\newcommand{\genericRVS}[1]{\mathbsf{#1}} 
\newcommand{\genericT}[2]{#1\left[#2\right]} 
\newcommand{\genericTC}[3]{#1_{#3}[#2]} 

\newcommand{\genericDFT}[1]{\MakeUppercase{#1}}
\newcommand{\genericIndex}[2]{#1_{#2}} 

\newcommand{\argmin}[1]{\arg \min_{#1}}

\newcommand{\card}[1]{|#1|}

\newcommand{\src}{x}
\newcommand{\sSrc}{\genericS{\src}}
\newcommand{\rvSrc}{\genericRV{\src}}

\newcommand{\rvSrcT}[1]{\genericT{\rvSrc}{#1}}

\newcommand{\rvFSrcT}[1]{\genericT{\genericRV{\genericDFT{\src}}}{#1}}

\newcommand{\rvsSrc}{\genericRVS{\src}}

\newcommand{\srcT}[1]{\genericT{\src}{#1}}
\newcommand{\quant}{\hat{\src}}
\newcommand{\sQuant}{\genericS{\hat{\src}}}
\newcommand{\rvQuant}{\genericRV{\quant}}
\newcommand{\rvsQuant}{\genericRVS{\quant}}

\newcommand{\rvQuantTC}[2]{\genericTC{\rvQuant}{#1}{#2}}
\newcommand{\rvQuantT}[1]{\genericT{\rvQuant}{#1}}

\newcommand{\quantT}[1]{\genericT{\quant}{#1}}

\newcommand{\sinfoEnc}{q}
\newcommand{\rvSinfoEnc}{\genericRV{\sinfoEnc}}
\newcommand{\rvsSinfoEnc}{\genericRVS{\sinfoEnc}}
\newcommand{\rvSinfoEncT}[1]{\genericT{\rvSinfoEnc}{#1}}
\newcommand{\sinfoEncT}[1]{\genericT{\sinfoEnc}{#1}}
\newcommand{\sSinfoEnc}{\genericS{\sinfoEnc}}

\newcommand{\sinfoDec}{w}

\newcommand{\rvsSinfoDec}{\genericRVS{\sinfoDec}}
\newcommand{\rvSinfoDecT}[1]{\genericT{\genericRV}{\sinfoDec}}

\newcommand{\srcAlph}{{\mathcal{\MakeUppercase{\src}}}}
\newcommand{\quantAlph}{{\mathcal{\MakeUppercase{\quant}}}}

\newcommand{\sinfoEncAlph}{{\mathcal{\MakeUppercase{\sinfoEnc}}}}

\newcommand{\encoder}{f}
\newcommand{\decoder}{g}

\newcommand{\dmin}{D_{\min}}

\newcommand{\reconFunc}{v}

\newcommand{\noiseVar}{z}
\newcommand{\rvNoiseVar}{\genericRV{\noiseVar}}

\newlength{\cheatLength}
\newlength{\secCheatLength}
\newcommand{\rdqboth}{R_{\mathrm{BOTH}}}
\newcommand{\rdqenc}{R_{\mathrm{ENC}}}
\newcommand{\rdqdec}{R_{\mathrm{DEC}}}
\newcommand{\rdqnone}{R_{\mathrm{NONE}}}
\newcommand{\rvu}{\genericRV{u}}
\begin{document}

\renewcommand{\textfraction}{0}

\renewcommand{\thefootnote}{} 

\title{Source Coding With Distortion Side Information At The Encoder}
\footnotetext{$^*$This work was conducted in part while R. Zamir was
  visiting the Digital Signal Processing Group at MIT.}

\author{
\begin{tabular}{cc}
\vspace{-0.07in}
\normalsize Emin Martinian, Gregory
W. Wornell&
\normalsize Ram Zamir$^*$\\ \vspace{-0.07in}
\normalsize Massachusetts Institute of Technology& \normalsize
Tel Aviv University\\ \vspace{-0.07in}
\normalsize Cambridge, MA~~USA& \normalsize Tel Aviv, Israel\\ \vspace{-0.07in}
\normalsize Email: \{emin,gww\}@allegro.mit.edu&
\normalsize Email: \{zamir\}@eng.tau.ac.il
\end{tabular}
}

\date{}
\maketitle
\thispagestyle{empty}
\pagestyle{empty}
\renewcommand{\thefootnote}{\arabic{footnote}} 

\normalsize

\begin{center}
\resizebox{\textwidth}{!}{%
\begin{minipage}{6.75in}
{\bf Abstract:} We consider lossy source coding when side information
affecting the distortion measure may be available at the encoder,
decoder, both, or neither.  For example, such distortion side
information can model reliabilities for noisy measurements, sensor
calibration information, or perceptual effects like masking and
sensitivity to context.  When the distortion side information is
statistically independent of the source, we show that in many cases
(\eg, for additive or multiplicative distortion side information)
there is no penalty for knowing the side information only at the
encoder, and there is no advantage to knowing it at the decoder.
Furthermore, for quadratic distortion measures scaled by the
distortion side information, we evaluate the penalty for lack of
encoder knowledge and show that it can be arbitrarily large.  In this
scenario, we also sketch transform based quantizers constructions
which efficiently exploit encoder side information in the
high-resolution limit.
\end{minipage}
}
\end{center}

\section{Introduction}

In many large systems such as sensor networks, communication networks,
and biological systems different parts of the system may each have
limited or imperfect information but must somehow cooperate.  Key
issues in such scenarios include the penalty incurred due to the lack
of shared information, possible approaches for combining information
from different sources, and the more general question of how different
kinds of information can be partitioned based on the role of each
system component.

One example of this scenario is when an observer records a signal
$\rvsSrc$ to be conveyed to a receiver who also has some additional
signal side information $\rvsSinfoDec$ which is correlated with
$\rvsSrc$.  As demonstrated by various researchers, in many cases the
observer and receiver can obtain the full benefit of the signal side
information even if it is known only by the receiver
\cite{it:Cover:2002} \cite{csiszar_korner} \cite{wynerZiv:76}.  

In this paper we consider a different scenario where instead the
observer has some distortion side information $\rvsSinfoEnc$ which
describes what components of the data are more sensitive to distortion
than others, but the receiver may not have access to $\rvsSinfoEnc$.
Specifically, let us model the differing importance of different
signal components by measuring the distortion between the $i$th source
sample, $\rvSrcT{i}$, and its quantized value, $\rvQuantT{i}$, by a
distortion function which depends on the side information
$\rvSinfoEncT{i}$: $d(\rvSrcT{i},\rvQuantT{i},\rvSinfoEncT{i})$.

In principle, one could treat the source-side information pair
$(\rvsSinfoEnc,\rvsSrc)$ as an ``effective composite source'', and
apply conventional techniques to quantize it.  Such an approach,
however, ignores the different effect $\rvsSinfoEnc$ and $\rvsSrc$
have on the distortion.  And as often happens in lossy compression,
good understanding of the distortion measure may lead to better
designs.
 
For example, a sensor may have side information corresponding to
reliability estimates for measured data (which may or may not be
available at the receiver).  This may occur if the sensor can
calibrate its accuracy to changing conditions (\eg, the amount of
light, background noise, or other interference present), if the sensor
averages data for a variety of measurements (\eg, combining results
from a number of sub-sensors) or if some external signal indicates
important events (\eg, an accelerometer indicating movement).

Alternatively, certain components of the signal may be more or less
sensitive to distortion due to masking effects or context
\cite{Jayant_1993}.  For example errors in audio samples following a
loud sound, or errors in pixels spatially or temporally near bright
spots are perceptually less relevant.  Similarly, accurately
preserving certain edges or textures in an image or human voices in
audio may be more important than preserving background
patterns/sounds.  Masking, sensitivity to context, etc., is usually a
complicated function of the entire signal.  Yet often
there is no need to explicitly convey information about this
function to the encoder.  Hence, from the point of
view of quantizing a given sample, it is reasonable to model such
effects as side information.

Clearly in performing data compression with distortion side
information, the encoder should weight matching the more important
data more than matching the less important data.  The importance of
exploiting the different sensitivities of the human perceptual system
are widely recognized by engineers involved in the construction and
evaluation of practical compression algorithms {\em when distortion
side information is available at both observer and receiver}.  In
contrast, the value and use of distortion side information known only
at either the encoder or decoder but not both has received relatively
little attention in the information theory and quantizer design
community.  The rate-distortion function with decoder-only side
information, relative to side information dependent distortion
measures (as an extension of the Wyner-Ziv setting
\cite{wynerZiv:76}), is given in \cite{csiszar_korner}.  A high
resolution approximation for this rate-distortion function for locally
quadratic weighted distortion measures is given in \cite{it:lzz:2000}.

We are not aware of an information-theoretic treatment of encoder-only
side information with such distortion measures.  In fact, the mistaken
notion that encoder only side information is never useful is common
folklore.  This may be due to a misunderstanding of Berger's result
that side information {\em which does not affect the distortion
measure} is never useful when known only at the encoder
\cite{berger71:book}.

In this paper we study the rate-distortion trade-off when side
information about the distortion sensitivity is available.  We show
that such distortion side information can provide an arbitrarily large
advantage (relative to no side information) even when the distortion
side information is known only at the encoder.  Furthermore, we show
that just as knowledge of signal side information is often only
required at the decoder, knowledge of distortion side information is
often only required at the encoder.  Beyond the theoretical results,
these observations serve as a useful guide for designing quantizers
with distortion side information.

We first illustrate how distortion side
information can be used even when known only by the observer with some
examples in \secref{sec:examples}.  Next, in
\secref{sec:problem-model}, we precisely define a problem model and
state the relevant rate-distortion trade-offs.  In
\secref{sec:main-results}, we present our main results characterizing
when knowledge of distortion side information is sufficient at only
the encoder and sketch one practical construction. 

\vspace{\secCheatLength}
\section{Examples}
\label{sec:examples}

\subsection{Discrete Uniform Source}
\label{sec:discr-unif-source}
Consider the case where the source, $\rvSrcT{i}$, corresponds to $n$
samples each uniformly and independently drawn from the finite
alphabet $\srcAlph$ with cardinality $\card{\srcAlph} \geq n$.  Let
$\rvSinfoEncT{i}$ correspond to $n$ binary variables indicating which
source samples are relevant.  Specifically, let the distortion measure
be of the form $d(\sinfoEnc,\src,\quant) = 0$ if and only if either
$\sinfoEnc=0$ or $\src=\quant$.  Finally, let the sequence
$\rvSinfoEncT{i}$ be statistically independent of the source with
$\rvSinfoEncT{i}$ drawn uniformly from the $n$ choose $k$ subsets with
exactly $k$ ones.

If the side information were unavailable or ignored, then losslessly
communicating the source would require exactly $n \cdot \log
\card{\srcAlph}$ bits.  A better (though still sub-optimal) approach
when encoder side information is available would be for the encoder to
first tell the decoder which samples are relevant and then send only
those samples.  This would require $n\cdot H_b(k/n) +
k\cdot\log\card{\srcAlph}$ bits where $H_b(\cdot)$ denotes the binary
entropy function.  Note that if the side information were also known at
the decoder, then the overhead required in telling the decoder which
samples are relevant could be avoided and the total rate required
would only be $k\cdot\log\card{\srcAlph}$.  We will show that this
overhead can in fact be avoided even without decoder side information.

Pretend that the source samples $\rvSrcT{0}$, $\rvSrcT{1}$, $\ldots$,
$\rvSrcT{n-1}$, are a codeword of an $(n,k)$ Reed-Solomon (RS) code (or
more generally any MDS%
\footnote{The desired MDS code always exists since we assumed
  $\card{\srcAlph} \geq n$.  For $\card{\srcAlph} < n$,
  near MDS codes exist which give asymptotically similar performance with an
  overhead that goes to zero as $n\rightarrow\infty$.}
code) with $\rvSinfoEncT{i}=0$ indicating an
erasure at sample $i$.  Use the RS {\em decoding} algorithm
to ``correct'' the erasures and determine the $k$ corresponding
information symbols which are sent to the receiver.  To reconstruct
the signal, the receiver {\em encodes} the 
$k$ information symbols using the encoder for the $(n,k)$ RS
code to produce the reconstruction $\rvQuantT{0}$, $\rvQuantT{1}$,
$\ldots$, $\rvQuantT{n-1}$.  Only symbols with $\rvSinfoEncT{i} = 0$
could have changed, hence $\rvQuantT{i} = \rvSrcT{i}$ whenever
$\rvSinfoEncT{i} = 1$ and the relevant samples are losslessly
communicated using only $k\cdot\log\card{\srcAlph}$ bits.

\begin{figure}[htb]
\begin{center}
\resizebox{6in}{1in}{\includegraphics{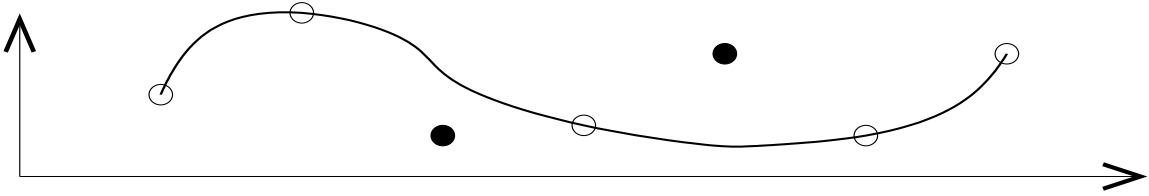}}
\caption{Losslessly encoding a source with $n=7$ points where only
$k=5$ points are relevant (\ie,
the unshaded ones), can be done by fitting a fourth degree curve to
the relevant points.  The resulting curve will require $k$ elements
(yielding a compression ratio of $k/n$) and will exactly reproduce the
desired points.\vspace{\cheatLength}
 \label{fig:rs_example}}
\end{center}
\end{figure}

As illustrated in \figref{fig:rs_example}, RS decoding
can be viewed as curve-fitting and RS encoding can be viewed as
interpolation.  Hence this source coding approach can be viewed as
fitting a curve of degree $k-1$ to the points of $\rvSrcT{i}$ where
$\rvSinfoEncT{i} = 1$.  The resulting curve can be specified using just
$k$ elements.  It perfectly reproduces $\rvSrcT{i}$ where $\rvSinfoEncT{i}
= 1$ and interpolates the remaining points.

\subsection{Gaussian Source}
\label{sec:gaussian-source}
A similar approach can be used to quantize a zero mean, unit variance,
complex Gaussian source relative to quadratic distortion using the
Discrete Fourier Transform (DFT).  Specifically, to encode the source
samples $\rvSrcT{0}$, $\rvSrcT{1}$, $\ldots$, $\rvSrcT{n-1}$, pretend
that they are samples of a complex, periodic, Gaussian, sequence with
period $n$, which is band-limited in the sense that only its first $k$
DFT coefficients are non-zero.  Using periodic, band-limited,
interpolation we can use only the $k$ samples for which 
$\rvSinfoEncT{i}=1$ to find the corresponding $k$ DFT coefficients,
$\rvFSrcT{0}$, $\rvFSrcT{1}$, $\ldots$, $\rvFSrcT{k-1}$.

The relationship between the $k$ relevant source samples and the $k$
interpolated DFT coefficients has a number of special properties.  In
particular this $k\times k$ transformation is unitary.  Hence, the DFT
coefficients are Gaussian with unit variance and zero mean.  Thus, the
$k$ DFT coefficients can be quantized with average distortion $D$ per
coefficient and $k \cdot R(D)$ bits where $R(D)$ represents the
rate-distortion trade-off for the quantizer.  To reconstruct the
signal, the decoder simply transforms the quantized DFT coefficients
back to the time domain.  Since the DFT coefficients and the relevant
source samples are related by a unitary transformation, the average
error per coefficient for these source samples is exactly $D$.

Note if the side information were unavailable or ignored, then at least
$n \cdot R(D)$ bits would be required.  If the side information were
losslessly sent to the decoder, then $n\cdot H_b(k/n) + k \cdot R(D)$ would
be required.  Finally, even if the decoder had knowledge of the side
information, at least $k\cdot R(D)$ bits would be needed.  Hence, the
DFT scheme achieves the same performance as when the side information
is available at both the encoder and decoder, and is strictly better
than ignoring the side information or losslessly communicating it.

\vspace{\secCheatLength}
\section{Problem Model}
\label{sec:problem-model}

Vectors and sequences are denoted in bold (\eg, $\genericS{x}$) with
the $i$th element denoted as $\genericT{x}{i}$.
Random variables are denoted using the sans serif
font (\eg, $\genericRV{x}$) while random vectors and sequences are
denoted with bold sans serif (\eg, $\genericRVS{x}$).  We denote
mutual information, entropy, and expectation as
$I(\genericRV{x};\genericRV{y})$, $H(\genericRV{x})$, 
$E[\genericRV{x}]$.  Calligraphic letters denote sets (\eg, $\src \in
\srcAlph$). 

We are primarily interested in a particular type of side information (which we
call ``distortion side information'') that is statistically independent of the
source but affects the distortion measure.  Specifically, we consider
the source coding with distortion side information problem defined as the tuple
\begin{equation}
(\srcAlph, \quantAlph, \sinfoEncAlph,
  p_{\rvSrc}(\src), p_{\rvSinfoEnc}(\sinfoEnc),
  d(\cdot,\cdot,\cdot)).
\end{equation}
A source $\rvsSrc$ consists of the $n$ samples $\rvSrcT{1}$, 
$\rvSrcT{2}$, $\ldots$, $\rvSrcT{n}$ drawn from the alphabet
$\srcAlph$.  The distortion 
side information $\rvsSinfoEnc$ likewise consists of $n$ samples drawn
from the alphabet $\sinfoEncAlph$.
These random variables are generated according to the distribution
\[
p_{\rvsSrc,\rvsSinfoEnc}(\sSrc,\sSinfoEnc) =
\prod_{i=1}^n
p_{\rvSrc}(\srcT{i}) \cdot
p_{\rvSinfoEnc}(\sinfoEncT{i}).
\]

A rate $R$ encoder, $\encoder(\cdot)$, maps a source as well as
possible side information to an index $i \in \{1,2,\ldots,2^{nR}\}$.
The corresponding decoder, $\decoder(\cdot)$, maps the resulting index
as well as possible decoder side information to a reconstruction of
the source.  Distortion for a source $\sSrc$ which is quantized and
reconstructed to the sequence $\sQuant$ taking values in the alphabet
$\quantAlph$ is measured via
\begin{equation}
d(\sSrc,\sQuant,\sSinfoEnc) = \frac{1}{n}\sum_{i=1}^n
d(\srcT{i},\quantT{i},\sinfoEncT{i}). 
\end{equation}
As usual, the rate-distortion function is the minimum rate such that
there exists a system where the distortion is at most $D$ with
probability approaching 1 as $n\rightarrow\infty$.  

The four scenarios where $\rvsSinfoEnc$ is available at the encoder,
decoder, both, or neither are illustrated in \figref{fig:si} along
with the symbol denoting each rate-distortion function.
\begin{prop}
The rate-distortion functions for the scenarios in \figref{fig:si} are
\begin{subequations}
\begin{align}
\label{eq:rd-none}
\rdqnone(D) &= \inf_{p_{\rvQuant|\rvSrc}(\quant|\src):
  E[d(\rvSrc,\rvQuant,\rvSinfoEnc)] \leq D} I(\rvSrc;\rvQuant)\\
\label{eq:rd-dec}
\rdqdec(D) &=  \inf_{p_{\rvu|\rvSrc}(u|\src), \reconFunc(\cdot,\cdot):
  E[d(\rvSrc,v(\rvu,\rvSinfoEnc),\rvSinfoEnc)] \leq D}
  I(\rvSrc;\rvu)-I(\rvu;\rvSinfoEnc)\\
\label{eq:rd-enc}
\rdqenc(D) &=  \inf_{p_{\rvQuant|\rvSrc,\rvSinfoEnc}(\quant|\src,\sinfoEnc):
  E[d(\rvSrc,\rvQuant,\rvSinfoEnc)] \leq D}
  I(\rvSrc,\rvSinfoEnc;\rvQuant) =   I(\rvSrc;\rvQuant|\rvSinfoEnc) +
  I(\rvQuant;\rvSinfoEnc)\\
\label{eq:rd-both}
\rdqboth(D) &= \inf_{p_{\rvQuant|\rvSrc,\rvSinfoEnc}(\quant|\src,\sinfoEnc):
  E[d(\rvSrc,\rvQuant,\rvSinfoEnc)] \leq D}
  I(\rvSrc;\rvQuant|\rvSinfoEnc).
\end{align}
\end{subequations}
\end{prop}
The rate-distortion functions in \eqref{eq:rd-none},
 \eqref{eq:rd-dec}, and \eqref{eq:rd-both} follow from standard results
 (\eg, \cite{berger71:book} \cite{it:Cover:2002} \cite{csiszar_korner}
 \cite{Gray_1973} \cite{wynerZiv:76}).  To obtain \eqref{eq:rd-enc} we
 can apply the classical 
 rate-distortion theorem to the ``super source'' $\rvsSrc' =
 (\rvsSrc,\rvsSinfoEnc)$ as suggested by Berger \cite{berger:private_comm}.
In the sequel we characterize the penalty or rate-loss
incurred by having side information available only at the encoder,
only at the decoder, or neither compared to full side information.

\newcommand{\sidiagram}[5]{%
\begin{minipage}{3in}
\begin{center}
\psfrag{X}{\Large $\rvsSrc$}
\psfrag{Y}{\Large $\rvsQuant$}
\psfrag{S}{\Large $\rvsSinfoEnc$}
\psfrag{W}{\Large $\rvsSinfoDec$}
\psfrag{ii}{\Large $i$}
\psfrag{Enc}{\Large#1}
\psfrag{Dec}{\Large#2}
\subfigure[#4 \label{#5}]{\resizebox{3in}{!}{\includegraphics{figs/#3}}} 
\end{center}
\end{minipage}
}

\begin{figure}[hbtp]
\begin{tabular}{lr}
\sidiagram{$\encoder(\rvsSrc)$}{$\decoder(i)
  $}{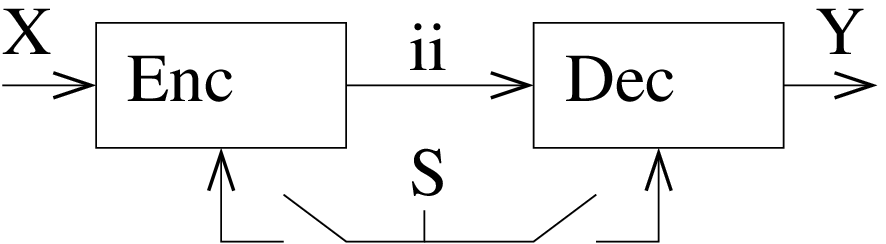}{$\rdqnone(D)$.}{fig:si:q_none}&    
\sidiagram{$\encoder(\rvsSrc)$}{$
 \decoder(i,\rvsSinfoEnc)$}{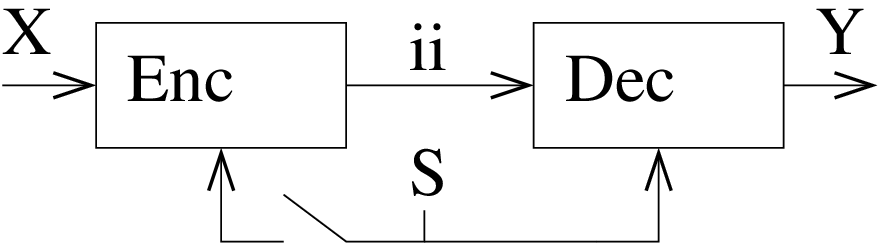}{$\rdqdec(D)$.}{fig:si:q_dec}\\  
\sidiagram{$\encoder(\rvsSrc,\rvsSinfoEnc)$}{$
 \decoder(i)$}{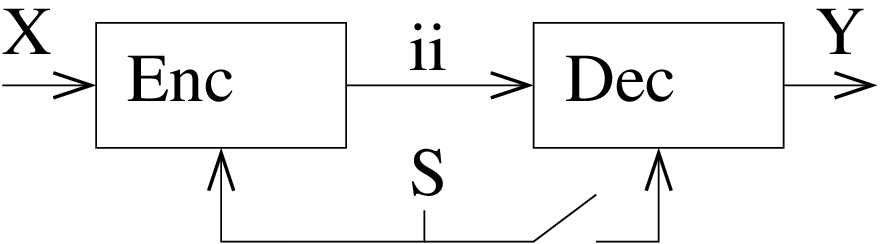}{$\rdqenc(D)$.}{fig:si:q_enc}&
\sidiagram{$\encoder(\rvsSrc,\rvsSinfoEnc)$}{$
 \decoder(i,\rvsSinfoEnc)$}{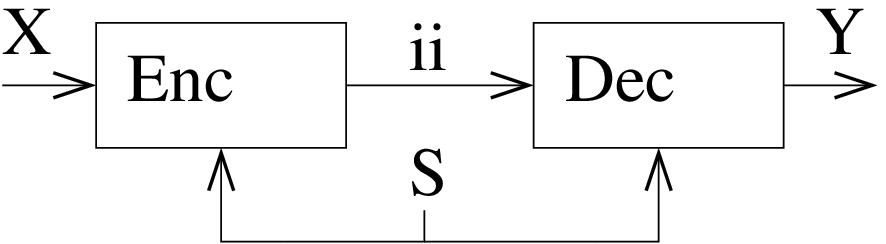}{$\rdqboth(D)$.}{fig:si:q_both}\\  
\end{tabular}
\caption{Possible scenarios and rate-distortion functions with
  distortion side information at the decoder (b), encoder (c),
  both (d), and neither (a).  The terms in (b) and
  (d) are also known as the Wyner-Ziv and conditional rate-distortion
  functions. \label{fig:si}}  
\end{figure}

\newcommand{\rvsNoiseVar}{\genericRVS{\noiseVar}}

\vspace{\secCheatLength}
\section{Main Results}
\label{sec:main-results}

A system with encoder only side information corresponds to a system
with a fixed codebook but  a variable partition which depends upon
$\rvsSinfoEnc$.\footnote{This structure also appears in the study of
robust codebooks \cite{itw:Zamir;2002}.}  As an almost trivial
example, consider an encoder which observes $\rvsSrc = \rvsNoiseVar
+\rvsSinfoEnc$ where $\rvsNoiseVar$ represents the true signal and
$\rvsSinfoEnc$ represents observation noise, \ie, $d(\noiseVar-\quant) =
d(\src-\sinfoEnc-\quant) = d'(\src,\quant,\sinfoEnc)$.  By shifting
the partition by $\rvsSinfoEnc$ to quantize $\rvsSrc-\rvsSinfoEnc$,
the encoder achieves optimal performance.  By contrast, systems with
decoder only side information correspond to fixed partitions with
variable codebooks and often can not exploit distortion side
information as easily.  In the following, we make these notions
precise for more general distortion measures.

\subsection{Rate-Distortion Trade-Offs}

We begin with the following theorems (proved in
\appref{app:proofs}) which show when side information at the
encoder can be optimally used even though such side information may be
useless if known only at the decoder.

\begin{thrm}
\label{th:gdiff:no-rate-loss}
Let distortion side information $\rvsSinfoEnc$ be statistically
independent of the  
source $\rvsSrc$ and let $\rvsSrc$ be uniformly distributed over a group
with distortion measured via $d(\src,\quant,\sinfoEnc) = d(\src
\ominus \quant, \sinfoEnc)$ 
where $\ominus$ represents a binary group operation.  Then the
rate-distortion function when $\rvsSinfoEnc$ is available at the
encoder is the same as when it is available at both encoder and
decoder, \ie, $\rdqenc(D) = \rdqboth(D)$.
\end{thrm}

To state a similar result for continuous sources, we require various
technical conditions describing a ``smooth'' source and distortion
measure.  Essentially, all that is required is that the source have a
density and finite differential entropy and that an entropy maximizing
distribution exists for the distortion measure of interest.  
For example, any vector source and distortion measure with
\begin{equation}
-\infty < h(\rvSrc) < \infty \textnormal{ and }
 E[||\rvSrc||^{\gamma_{\rvSinfoEnc}}] < \infty 
 \textnormal{ and }
d(\src,\quant,\sinfoEnc) = \alpha_{\sinfoEnc} + \beta_{\sinfoEnc} \cdot
||\src-\quant||^{\gamma_{\sinfoEnc}} \ \ \forall \sinfoEnc
\end{equation}
will satisfy the required conditions provided $\alpha_{\rvSinfoEnc}$,
$\beta_{\rvSinfoEnc}$, $\gamma_{\rvSinfoEnc}$ are non-negative.  See
\cite{Linder_1994} or \cite{our_preprint} for a more detailed
discussion of the necessary technical conditions.

\begin{thrm}
\label{th:hi-res:no-rate-loss}
Let $\rvsSinfoEnc$ be statistically independent of the source
$\rvsSrc$ and consider any ``smooth'' source and distortion measure
satisfying the conditions in \cite[Theorem 1]{Linder_1994} for each
$\sinfoEnc \in \sinfoEncAlph$.  Then the
rate-distortion function when 
$\rvsSinfoEnc$ is available only at the encoder is asymptotically the
same as when it is available at both encoder and decoder, \ie,
$\lim_{D \rightarrow \dmin} \rdqenc(D) - \rdqboth(D) =
0$.\footnote{Usually $\dmin=0$, but to allow for more general
  distortion measures we define $\dmin$ as the minimum achievable
  distortion when arbitrarily high rates are allowed.}
\end{thrm}

Finally, in addition to the previous theorems showing when only the
encoder requires $\rvsSinfoEnc$, we have the following result stating
when $\rvsSinfoEnc$ is useless to the decoder.
\begin{thrm}
\label{th:sinfo-useless-at-dec}
Let the distortion side information $\rvsSinfoEnc$ be statistically
independent of 
the source $\rvsSrc$ and consider scaled
distortion measures of the form
$d(\src,\quant,\sinfoEnc) = d_0(\sinfoEnc) \cdot d_1(\src,\quant)$.
Then the rate-distortion function for $\rvsSinfoEnc$ available at the
decoder is the same as when $\rvsSinfoEnc$ is available at neither
encoder nor decoder, \ie, $\rdqdec(D) = \rdqnone(D)$.
\end{thrm}

Combining our results shows that in many cases knowledge of
$\rvsSinfoEnc$ is optimal at the encoder and useless at the decoder.
\begin{corol}
\label{corol:combine}
For sources and side information weighted difference distortion
measures satisfying the conditions in
Theorems~\ref{th:gdiff:no-rate-loss} and \ref{th:sinfo-useless-at-dec}
(or respectively in Theorems~\ref{th:hi-res:no-rate-loss} and
\ref{th:sinfo-useless-at-dec}), 
$\rdqenc(D)-\rdqboth(D)=0$ and $\rdqnone(D)-\rdqdec(D)=0$, or respectively,
$\lim_{D\rightarrow\dmin} \rdqenc(D)-\rdqboth(D)=0$ and
$\lim_{D\rightarrow\dmin} \rdqnone(D)-\rdqdec(D)=0$.
\end{corol}

\subsection{The Penalty for Lack of Encoder Knowledge}

Consider generalizing the commonly used quadratic distortion model by
scaling the distortion as a function of the side information as in
\cite{it:lzz:2000}. Specifically, let
$d(\rvSinfoEnc,\rvSrc,\rvQuant) = \rvSinfoEnc \cdot (\rvSrc-\rvQuant)^2$.
For this scenario, \cite{it:lzz:2000} implies that $\rdqboth(D) =
h(\rvSrc) - (1/2) \ln(2 \pi e D) + (1/2)
E[\ln \rvSinfoEnc]$ while $\rdqdec(D) = h(\rvSrc) - (1/2) \ln(2 \pi e D)
+ (1/2) \ln E[\rvSinfoEnc]$. Combining this with
\corolref{corol:combine} shows that the
asymptotic penalty for lack of encoder knowledge of $\rvSinfoEnc$ is
$(1/2)\cdot(\ln E[\rvSinfoEnc] - E[\ln 
\rvSinfoEnc])$
nats per sample.  Table~\ref{tab:rate-gap} evaluates
this penalty for various distributions of $\rvSinfoEnc$.  Note that in
many cases, the rate loss can be made arbitrarily large by choosing
the appropriate shape parameter to place more probability near
$\rvSinfoEnc = 0$.  Intuitively, this occurs because when $\rvSinfoEnc
\approx 0$, the informed encoder can transmit almost zero rate while
the uninformed encoder must transmit a large rate to achieve high
resolution.  Furthermore, all but one of these distributions would
require infinite rate to losslessly communicate the side information.
\begin{table}[hbtp]
\caption{The rate-penalty (in nats) for not knowing side-information with the
  given distribution at the encoder.  Euler's constant is denoted by
  $\gamma$. \label{tab:rate-gap}}
\resizebox{\textwidth}{!}{%
\begin{minipage}{6.75in}
\newcommand{\rowend}{\textnormal{\begin{Huge}\phantom{A}\end{Huge}}\\\hline}
\[
\begin{array}{|l|l|l|}
\hline
\textnormal{Distribution Name}& \textnormal{Density for $\rvSinfoEnc$} &
\textnormal{Rate Gap in nats}\rowend 
\textnormal{Exponential} & \tau \exp(- \sinfoEnc \tau) & -\frac{1}{2} \ln
\gamma \approx 0.2748 \rowend 
\textnormal{Uniform} & 1_{\sinfoEnc \in [0,1]} & \frac{1}{2}(1-\ln 2) \approx
0.1534\rowend 
\textnormal{Lognormal} & \frac{1}{\sinfoEnc \sqrt{2\pi Q^2}} \exp \left[
  -\frac{(\ln \sinfoEnc - M)^2}{2 Q^2}\right] & \frac{Q^2}{4}\rowend 
\textnormal{Pareto} & \frac{a^b}{\sinfoEnc^{a+1}}, \sinfoEnc \geq b > 0, a >
1 &
\frac{1}{2}\left[ \ln \frac{a}{a-1} - 1/a\right]\rowend
\textnormal{Gamma} & \frac{b (b \sinfoEnc)^{a-1}
  \exp(-b \sinfoEnc)}{\Gamma(a)} & 
  \frac{1}{2}\left\{\ln a - \frac{d}{dx}
  [\ln\Gamma(x)]_{x=a}\right\}\approx \frac{1}{2a}\rowend
\textnormal{Pathological} & (1-\epsilon)\delta(\sinfoEnc-\epsilon) +
\epsilon\delta(\sinfoEnc-1/\epsilon)&
\frac{1}{2}\ln(1+\epsilon-\epsilon^2)-\frac{1-2\epsilon}{2}\ln
\epsilon \approx
\frac{1}{2}\ln \frac{1}{\epsilon}\rowend
\textnormal{Positive Cauchy} & \frac{2/\pi}{1+\sinfoEnc^2}, \sinfoEnc
\geq 0 & \infty\rowend 
\end{array}
\]
\end{minipage}}
\end{table}

\subsection{Quantizer Design}
\label{sec:quantizer-design}

\newcommand{\SEZero}{\genericIndex{\sinfoEnc}{0}}
\newcommand{\SEOne}{\genericIndex{\sinfoEnc}{1}}

As discussed in \secref{sec:examples}, for distortion side information
indicating that a given source sample is relevant or completely
irrelevant, a transform followed by a scalar
quantizer\footnote{Entropy coding the scalar quantizers output
is also possible without changing this result.}
efficiently exploits encoder side information.  To generalize this transform
coding construction, consider two-level side information with the
alphabet $\sinfoEncAlph = \{\SEZero, \SEOne\}$ where
$\SEOne \geq \SEZero \geq 0$ and distortion is measured via
$d(\rvSinfoEnc,\rvSrc,\rvQuant) = \rvSinfoEnc \cdot
(\rvSrc-\rvQuant)^2$.  Furthermore, let a random $k$ out of $n$
samples of $\rvsSinfoEnc$ take the value $\SEOne$ while the other
$n-k$ samples take the value $\SEZero$.  If $\rvsSinfoEnc$ is
known at both encoder and decoder then the optimal strategy is to use
a rate $R_0$ quantizer for samples when $\rvSinfoEnc=\SEZero$ and
a rate $R_1 \geq R_0$ quantizer when $\rvSinfoEnc=\SEOne$ such
that the overall rate or distortion constraint is satisfied.

\newcommand{\errSigTerm}{e}
\newcommand{\firstErrSigT}[1]{\genericT{\genericRV{\errSigTerm}}{#1}}
\newcommand{\firstErrQuantT}[1]{\genericT{\hat{\genericRV{\errSigTerm}}}{#1}}
\newcommand{\FirstErrSigT}[1]{\genericT{\genericRV{\genericDFT{\errSigTerm}}}{#1}}
\newcommand{\secErrSigT}[1]{\genericT{\genericRV{\errSigTerm'}}{#1}}

To asymptotically achieve the same performance via transform coding
when $\rvsSinfoEnc$ is known only at the encoder, we can use the
following procedure.  First, quantize the $k$ more important source
samples where $\rvSinfoEncT{i} = \SEOne$ with a rate $R_0$
quantizer to produce $\rvsQuant_1$.  Define the first stage error
signal as $\firstErrSigT{i} = \rvSrcT{i}-\rvQuantTC{i}{1}$ where we
assume $\rvQuantTC{i}{1}=0$ when $\rvSinfoEncT{i}=\SEZero$ since
these less important samples have not yet been quantized. Next use
band-limited interpolation to find the $k$ DFT coefficients
$\FirstErrSigT{i}$ such that the IDFT of $\FirstErrSigT{i}$ accurately
reproduces $\firstErrSigT{i}$ when $\rvSinfoEncT{i}=\SEOne$.
Quantize these coefficients using a rate $R_1-R_0$ quantizer.  Define
the second stage error signal as $\secErrSigT{i} =
\rvSrcT{i}-\rvQuantTC{i}{1}-\firstErrQuantT{i}$ where
$\firstErrQuantT{i}$ represents the IDFT of the quantized
$\FirstErrSigT{i}$.  Finally, quantize the $n-k$ samples of $\secErrSigT{i}$
where $\rvSinfoEncT{i} = \SEZero$ using a rate $R_0$ quantizer to
produce $\rvsQuant_2$.

\begin{figure}[hbt]
\begin{center}
\begin{tabular}{lr}
\resizebox{2.05in}{!}{\includegraphics{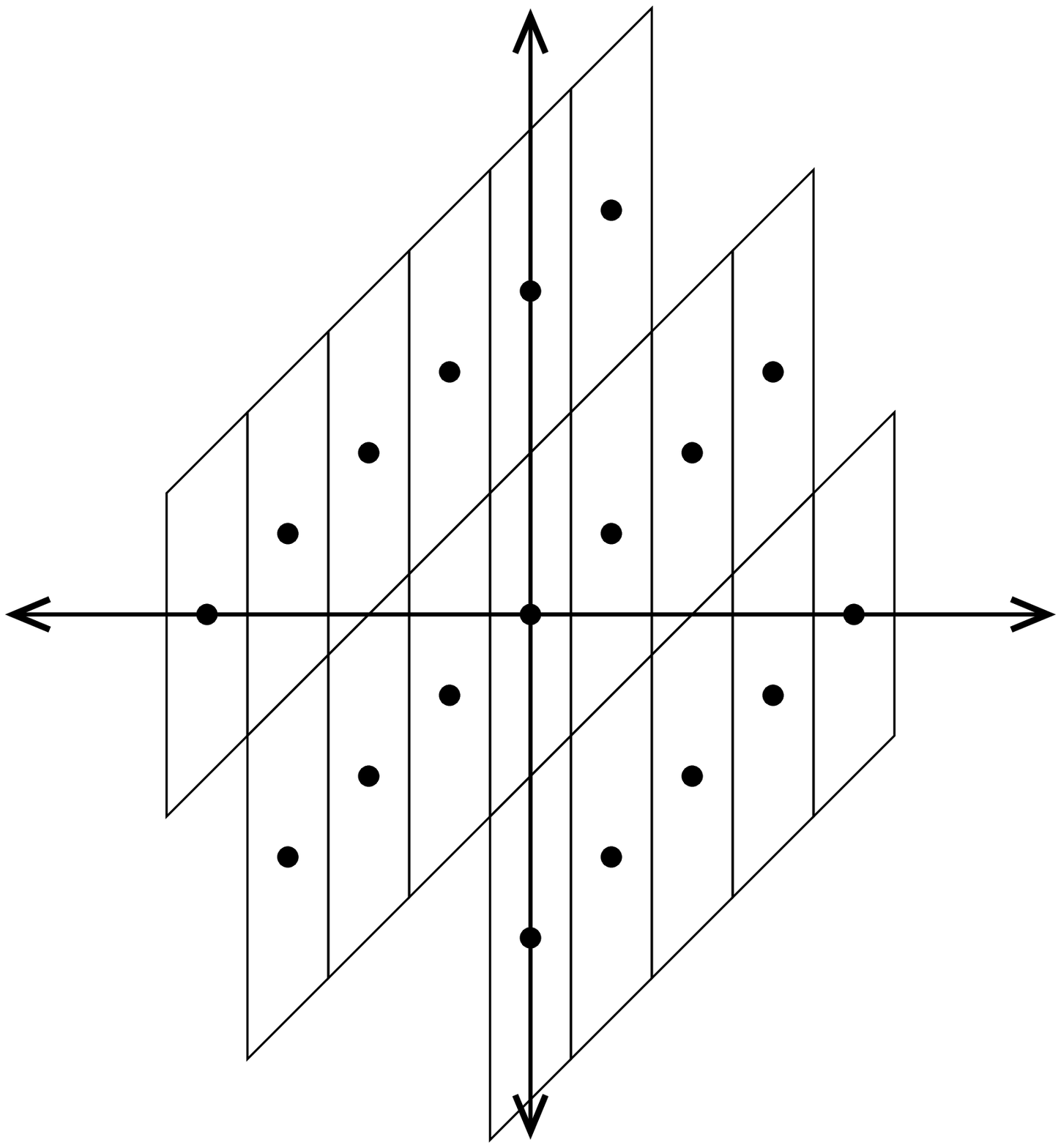}}
\hspace{.25in} &
\hspace{.25in}
\resizebox{2.05in}{!}{\includegraphics{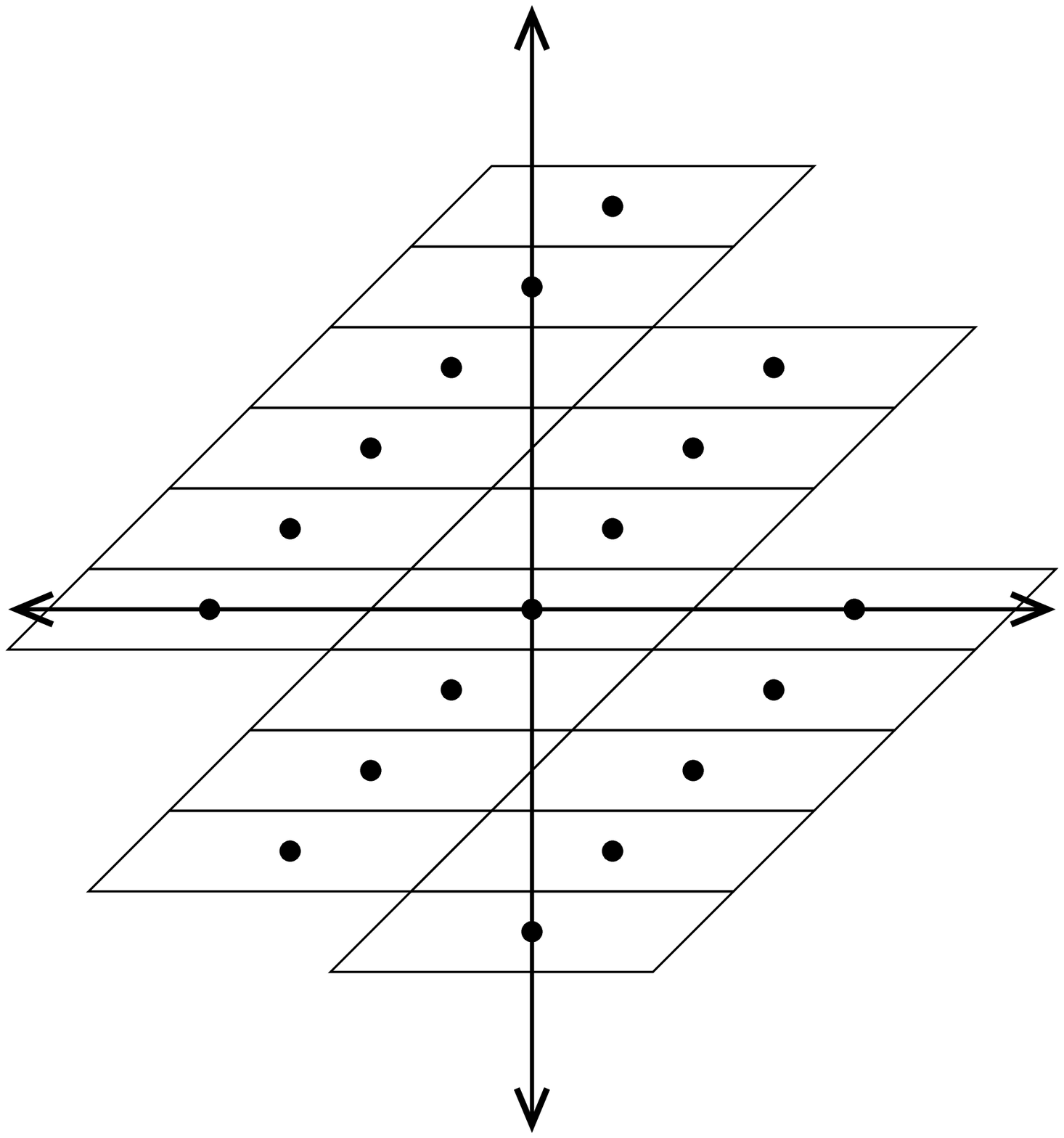}}
\end{tabular}
\caption{The quantization points and possible partitions for a
  transform coder.  If the encoder knows the horizontal error
  (respectively, vertical error) is more important, it can use the
  partition on the left to increase horizontal accuracy
  (resp., vertical accuracy).  The decoder only needs to know the
  quantization point not the
  partition. \vspace{\cheatLength}\vspace{\cheatLength} 
   \label{fig:rot_q}}
\end{center}
\end{figure}

The receiver obtains the reconstruction $\rvQuantT{i} =
\rvQuantTC{i}{1} + \rvQuantTC{i}{2} + \firstErrQuantT{i}$ consisting
of a rate $R_0$ scalar quantization of each source sample and a
quantized shift $\firstErrQuantT{i}$.  Although the receiver can not
deduce from $\firstErrQuantT{i}$ which samples were more important,
$\firstErrQuantT{i}$ was chosen by the encoder to make the
quantization of the more important samples more accurate.  As
illustrated in \figref{fig:rot_q} for $(n,k)=(2,1)$, this type of
system corresponds to a quantization lattice where the encoder can choose the
partition to shape the error based on the side information.  It is
possible to show that in high resolution this system approaches the
performance of a fully informed system (\ie, using a rate $R_0$
quantizer when $\rvSinfoEncT{i}=\SEZero$ and a rate $R_1$
quantizer when $\rvSinfoEncT{i}=\SEOne$) \cite{our_preprint}.
Conceptually, in the high resolution limit, edge effects become
negligible and the shape of each cell in \figref{fig:rot_q} approaches
a rectangle.  This system specializes to the one in
\secref{sec:gaussian-source} when $\SEZero=0$ and can be further
generalized to larger side information alphabets \cite{our_preprint}.

\appendix

\vspace{\secCheatLength}
\section{Proofs}
\label{app:proofs}

\begin{proof}[Proof of \thrmref{th:gdiff:no-rate-loss}:]
For a finite group, choosing $\rvNoiseVar^*$ to maximize
$H(\rvNoiseVar|\rvSinfoEnc)$ subject to the constraint
$E[d(\rvNoiseVar,\rvSinfoEnc)] \leq D$ yields the following lower bound on
$\rdqenc(D)$: 
\begin{align}
I(\rvQuant;\rvSrc,\rvSinfoEnc) &= H(\rvSrc) + H(\rvSinfoEnc) -
H(\rvSrc,\rvSinfoEnc|\rvQuant)\\
&= \log |\srcAlph| + H(\rvSinfoEnc) -
H(\rvSinfoEnc|\rvQuant) - 
H(\rvQuant-\rvSrc|\rvSrc,\rvSinfoEnc)\\
\label{eq:gdiff:nrl:cond-ret-ent}
&\geq \log |\srcAlph|  - 
H(\rvQuant-\rvSrc|\rvSinfoEnc)\\
&\geq
\log |\srcAlph| - H(\rvNoiseVar^*|\rvSinfoEnc)
\end{align}
where \eqref{eq:gdiff:nrl:cond-ret-ent} follows since conditioning
reduces entropy.
Choosing the test-channel distribution $\rvQuant = \rvNoiseVar^* + \rvSrc$
achieves this bound with equality and must therefore be optimal.
Furthermore, since $\rvQuant$ and $\rvSinfoEnc$ are statistically
independent for 
this test-channel distribution, $I(\rvQuant;\rvSinfoEnc) = 0$ and thus
comparing \eqref{eq:rd-enc} and to \eqref{eq:rd-both} shows
$\rdqenc(D) = \rdqboth(D)$ for finite groups.  The same argument holds
for continuous groups with entropy replaced by differential entropy
and $|\srcAlph|$ replaced by the Lebesgue measure of $\srcAlph$.  For
more general groups (\eg, mixed groups with both discrete and
continuous components), a more complicated convexity argument is
required \cite{our_preprint}.
\end{proof}

\begin{proof}[Proof Sketch For \thrmref{th:hi-res:no-rate-loss}:]
Due to space constraints we only sketch the ideas behind the proof.
As for \thrmref{th:gdiff:no-rate-loss}, we can develop a lower bound for
$\rdqboth(D)$ using an entropy maximizing distribution and the Shannon
lower bound \cite{Linder_1994}.  Then by using the resulting
test-channel distribution for $\rdqenc(D)$ we can show that
$I(\rvQuant;\rvSinfoEnc)$ goes to zero in the high resolution limit
and therefore $\rdqenc(D)\rightarrow\rdqboth(D)$.
\end{proof}

\begin{proof}[Proof of \thrmref{th:sinfo-useless-at-dec}:]
When side information is available only at the decoder, Wyner-Ziv
coding is optimal \cite{wynerZiv:76}.  First we compute the optimal
reconstruction function $\reconFunc(\cdot,\cdot)$:
\begin{align}
\reconFunc(u,\sinfoEnc) &= \argmin{\quant}
E[d(\quant,\rvSrc,\rvSinfoEnc)|\rvSinfoEnc=\sinfoEnc,\rvu = u]\\
\label{eq:scaled-dist}
&= \argmin{\quant}
d_0(\sinfoEnc)E[d_1(\quant,\rvSrc)|\rvSinfoEnc=\sinfoEnc,\rvu = u]\\
\label{eq:scaling-irrelevent}
&= \argmin{\quant} E[d_1(\quant,\rvSrc)|\rvSinfoEnc=\sinfoEnc,\rvu =
  u]\\
\label{eq:sinfo-irrelevent}
&= \argmin{\quant} E[d_1(\quant,\rvSrc)|\rvu=u]
\end{align}
where \eqref{eq:scaled-dist} follows by the assumption that we have a
separable distortion measure and \eqref{eq:sinfo-irrelevent} follows
because $\rvSinfoEnc$ is statistically independent of $\rvSrc$ (by
assumption) and independent of $\rvu$ (since $\rvu$ is generated at
the encoder from $\rvSrc$).  Thus since neither the optimal
reconstruction function, $\reconFunc(\cdot,\cdot)$ nor the auxiliary
random variable, $\rvu$, depend on $\rvSinfoEnc$, knowing
$\rvSinfoEnc$ at only the decoder provides no advantage.
\end{proof}

\small
\bibliographystyle{ieeetr}
\bibliography{paper}

\end{document}